\documentclass[iicol]{sn-jnl}

\usepackage{graphicx}%
\usepackage{multirow}%
\usepackage{amsmath,amssymb,amsfonts}%
\usepackage{amsthm}%
\usepackage{mathrsfs}%
\usepackage[title]{appendix}%
\usepackage{xcolor}%
\usepackage{textcomp}%
\usepackage{manyfoot}%
\usepackage{booktabs}%
\usepackage{algorithm}%
\usepackage{algorithmicx}%
\usepackage{algpseudocode}%
\usepackage{listings}%
\usepackage[english]{babel}
\theoremstyle{thmstyleone}%
\theoremstyle{thmstyletwo}%

\theoremstyle{thmstylethree}%

\begin{document}

\title[Article Title]{Hydrodynamical Misner–Sharp Formulation for Gravitational Collapse in Scalar–Tensor Theories}


\author[1]{\fnm{Jose A.R. } \sur{Cembranos}}\email{cembra@ucm.es; \textcolor{black}{ORCID: 0000-0002-4526-7396}}

\author*[1]{\fnm{Luis} \sur{Diaz-Gimenez}}\email{luisdi07@ucm.es; \textcolor{black}{ORCID: 0009-0004-6406-7383}}

\affil[1]{\orgdiv{Departamento de F\'isica Te\'orica and IPARCOS}, \orgname{Facultad de Ciencias F\'isicas, Universidad Complutense de Madrid}, \orgaddress{\city{Madrid}, \postcode{28040}, \country{Spain}}}


\abstract{
We study the dynamics of gravitational collapse within the framework of scalar-tensor theories of gravity. By adopting the Einstein frame, we develop a novel hydrodynamical formulation inspired by the Misner-Sharp approach, which allows for a self-consistent description of the interaction between the fluid and the scalar field. We derive the generalized Tolman–Oppenheimer–Volkoff (TOV) equation for hydrostatic equilibrium, identifying how the scalar field modifies the balance between pressure gradients and gravitational attraction through an effective mass, pressure and a non-minimal coupling term. This equilibrium condition is subsequently used to construct consistent initial configurations for the collapse process. To illustrate the dynamics, we analyze the initial phases of the contraction and the junction conditions required to match the interior hydrodynamical solution with a suitable exterior vacuum within this limit. Our results show that the scalar field introduces a dynamical feedback that accelerates the collapse compared to General Relativity. This formulation provides a robust theoretical framework for studying strong-field phenomena and potential black hole formation in alternative gravity theories.}




\maketitle

\section{Introduction}\label{sec1}

The phenomenon of gravitational collapse lies at the very heart of modern astrophysics and General Relativity (GR), describing the inexorable contraction of a massive body under its own gravitational attraction. Beginning with the pioneering work of Oppenheimer and Snyder in 1939 \cite{OS39}, who formulated the first exact model of a pressureless dust star collapsing in a Schwarzschild exterior, gravitational collapse has provided both a theoretical laboratory for strong‐field gravity and a window into the formation of compact objects such as neutron stars and black holes. In its simplest form, the collapse is governed by the Einstein field equations coupled to an appropriate equation of state; however when one considers modified theories of gravity, additional degrees of freedom may play an important role in strong field astrophysical settings.

Scalar‑tensor theories of gravity extend GR by introducing one or more scalar degrees of freedom which couple directly to the metric and to matter. The original Brans–Dicke theory \cite{BD61}, motivated by Mach’s principle, posits a single scalar field $\phi$ whose inverse determines the effective gravitational constant and which interacts with the stress-energy tensor of the material fields via a coupling function. More general constructions allow for arbitrary scalar potentials $V(\phi)$ and more complicated couplings. In these theories, it is possible to perform a conformal transformation to work in what is known as the Einstein frame of the theory. In this frame the Einstein equations remain unaltered and the theory can be interpreted as simply as adding a scalar field coupled to matter to standard GR. Such theories arise naturally in low‑energy limits of string theory \cite{Callan85,Damour94} and allow the construction of cosmological inflation models without fine-tuning of the potential \cite{La89,Steindhardt90}.\par

Interest in scalar-tensor theories decreased drastically when observational data showed that the coupling parameter had to be strongly suppressed to match the data, making the theory predict results almost identical to those of GR. However, the theory regained its popularity after it was proven that, since the coupling is not necessarily a fixed parameter, some cosmological processes could be responsible for making the said coupling small \cite{Damour93}. Consequently, the effects of scalar-tensor gravity would only be observable in processes involving the extreme gravity regime. This highlights the importance of studying stellar collapse in these theories. Crucially, the dynamics of the scalar field could alter both the internal evolution of a collapsing star and the structure of the exterior spacetime, leading to potentially significant and observable deviations from the predictions of GR.

In this work, we undertake a systematic study of spherically symmetric Lagrangian gravitational collapse in a general scalar-tensor theory. By formulating the problem in the Einstein frame, we derive a set of modified Misner–Sharp equations, which govern the evolution of the metric, the fluid and the scalar field together with a barotropic equation of state. New terms proportional to the scalar coupling and to the pressure and density of the new spin-0 mediator appear in the hydrodynamic equations; these represent both a direct new force on the fluid and a source of additional effective mass and pressure from the scalar sector.

A key aspect of any collapse model is the specification of initial data in hydrostatic equilibrium, commonly obtained by solving the Tolman–Oppenheimer–Volkoff (TOV) equation. In Sec.3 we derive the modified TOV condition, which now contains contributions from the scalar energy density and pressure. The resulting equation generalizes the classical balance between pressure gradients and gravitational attraction by incorporating source terms, thereby determining how the scalar field alters the maximum mass, radius, and stability of compact configurations.

To build intuition and provide test cases for numerical evolution, Sec.4 constructs the simplest nontrivial solution: a homogeneous interior in which both the fluid and scalar field depend only on time. Here, the metric reduces to a Friedmann-Lemaître-Robertson-Walker (FLRW) form, and the Klein–Gordon equation acquires an effective friction term from the Hubble expansion plus a source term from the matter coupling. Specializing in Brans–Dicke theory with dust, we obtain analytic expressions for the energy density and perturbative solutions for the spin-0 field, demonstrating that the scalar energy density can grow more rapidly than the fluid energy density as collapse proceeds. Numerical illustrations show how stronger coupling accelerates collapse and drives the scalar field to dominate.

In addition to the interior, an astrophysical realistic collapse must join onto an asymptotically flat vacuum exterior. In Sec.5 we propose the Janis–Newman–Winicour (JNW) metric, a static spherically symmetric vacuum solution of massless scalar‐tensor gravity in the Einstein frame. We discuss why, despite not taking into account the gravitational scalar radiation emitted in the collapse, we regard this metric as a good approximation during the first instants of the collapse. 

Sec.6 outlines the conditions for a smooth metric match at the stellar surface, requiring continuity of the first and second fundamental forms. The conditions the scalar field must satisfy are also introduced and studied. The relations between the inner and outer metric parameters are found and the match is shown to be viable under certain proper assumptions.

Finally, in Sec.7 we discuss the implications of our findings. The inclusion of scalar interactions introduces qualitatively new behavior in the collapse process: additional forces, modified equilibrium configurations, and the emission of gravitational radiation even in spherically symmetric situations. These effects may have observational signatures in gravitational waveforms and fundamentally change the final state of the collapse. Moreover, our formalism is readily extendable to include different couplings, nontrivial potentials, or more complex equations of state; opening avenues for future work on black hole formation and astrophysical tests of modified gravity.

In summary, this work provides a comprehensive analytical and computational framework for studying gravitational collapse in scalar‑tensor gravity, combining modified hydrodynamics, equilibrium solutions, interior‐exterior matching, and illustrative examples to elucidate the role of scalar interactions in strong‑field astrophysics.

Unless explicitly stated otherwise, we will work with units in which $G=c=1$ and using the signature convention $(-,+,+,+)$ for the metric.

\section{Hydrodynamic equations}\label{sec2}

The purpose of this section is to derive a Lagrangian hydrodynamic formulation for a spherically symmetric collapse in scalar-tensor gravity similar to that of Misner-Sharp \cite{Misner-Sharp} in standard GR. Nariai and Matsuda previously conducted similar work \cite{Nariai72,Matsuda73} but they worked in the Jordan frame, which we believe makes the resulting equations considerably harder to interpret.\par 
We begin by establishing the action, which in the Einstein frame looks like
\begin{align}
    S =&  \int d^4 x \sqrt{-g} \left(\frac{1}{16\pi} R - \frac{1}{2} g^{\mu\nu} \partial_\mu \phi \partial_\nu \phi -V(\phi) \right) \nonumber \\
    &+ S_{\text{m}} \left[ \Psi_{\text{m}}, a^2(\phi) g_{\mu\nu} \right],
\end{align}
where $S_m$ is the action corresponding to the matter content, that we will treat like a perfect fluid, and \(a^2(\phi)\) is the conformal factor that takes the metric from the Einstein frame ($g_{\mu\nu}$) to the Jordan frame ($\tilde{g}_{\mu\nu}$),
\begin{equation}
    \tilde{g}_{\mu\nu} = a^2(\phi) g_{\mu\nu}.
\end{equation}
By varying \(S\) one obtains the equations of motion for the metric and the scalar field:
\begin{align}
     R_{\mu\nu} - \frac{1}{2} g_{\mu\nu} R = 8\pi T_{\phi \mu\nu} + 8\pi T_{\mu\nu},\label{eq_Einstein} \\
     \Box \phi = - \alpha(\phi) T + \frac{\partial V(\phi)}{\partial \phi},\label{eq_KG} 
\end{align}
with \(\alpha(\phi) \equiv \frac{\partial \ln a(\phi)}{\partial \phi}\) and where  $T_{\phi\mu\nu}$ refers to the stress-energy tensor for the scalar field,
\begin{equation}
    T_{\phi \mu\nu} = \partial_\mu \phi \partial_\nu \phi - \frac{1}{2} g_{\mu\nu} g^{\alpha\beta} \partial_\alpha \phi \partial_\beta \phi - V(\phi) g_{\mu\nu}.
\end{equation}
On the other hand, the tensor \(T_{\mu\nu}\) refers to the stress-energy of a perfect fluid and its interaction with the scalar field. We will assume it reads
\begin{align}
T_{\mu\nu} =({\rho} +{p}) {u}_\mu {u}_\nu + {p} {g}_{\mu\nu}.
\end{align}
Note that the quantities \( {\rho}   \) and \( {p} \) are the effective density and pressure in the Einstein frame. Do not confuse with the same quantities in the Jordan frame, where they are directly related to physical properties of the matter content. The trace of this tensor is simply $T=-\rho+3p$ and it appears in the coupling term of equation (\ref{eq_KG}).

By assuming spherical symmetry, the metric can be written without loss of generality as
\begin{equation}
    ds^2 = -e^{2\Phi(\mu,t_c)} dt_c^2 + \frac{r'^2}{\Gamma^2(\mu,t_c)}d\mu^2 + r^2(\mu,t_c) d\Omega^2,
\end{equation}
where \( t_c \) is the proper time for an observer moving with the fluid, and $\mu$ can be thought of as the mass trapped within a Schwarzschild radius \( r \). The functions $\Phi(\mu,t_c)$ and $\Gamma(\mu,t_c)$ will be determined later from equation (\ref{eq_Einstein}). In this situation, the stress-energy tensor for the scalar field may then take the form
\begin{equation}
    T_{\phi\mu\nu} = 
    \left(\begin{smallmatrix}
    (\dot{\phi}^2 - p_\phi) e^{2\Phi} & e^{\Phi} \dot{\phi} \phi' & 0 & 0 \\
    e^{\Phi} \dot{\phi} \phi' & {\phi'}^{2} + p_\phi \frac{r'^2}{\Gamma^2} & 0 & 0 \\
    0 & 0 & p_\phi r^2 & 0 \\
    0 & 0 & 0 & p_\phi r^2 \sin^2 \theta
    \end{smallmatrix}\right),
\end{equation}
where the following notation for the derivatives is being used:
\[
\dot{A} = \frac{1}{e^\Phi} \frac{\partial A}{\partial t_c}, \quad A' = \frac{\partial A}{\partial \mu},
\]
where $A$ is a generic funtion. We have also defined the pressure of the scalar field as
\begin{equation}
    p_\phi = \frac{1}{2} \left( \dot{\phi}^2 - \phi'^2 \frac{\Gamma^2}{r'^2} \right) - V(\phi).
\end{equation}
In the reference system we are working on, the 4-velocity of the fluid will be \( u^\mu = e^{-\Phi} \vec{e}_t \) and, therefore, we may write the matter stress-energy tensor as
\begin{equation}
   T_{\mu\nu} = 
\begin{pmatrix}
\rho e^{2\Phi} & 0 & 0 & 0 \\
0 & {p} r'^2 /\Gamma^{2} & 0 & 0 \\
0 & 0 & {p} r^{2} & 0 \\
0 & 0 & 0 & {p} r^{2} \sin^{2} \theta
\end{pmatrix}. 
\end{equation}

Our aim is now to deduce the hydrodynamic equations of this system. We start by considering that the particle number is preserved as in GR,
\begin{equation}
    \nabla_\mu (n u^\mu) = 0 \Rightarrow \frac{\dot{(nr^2)}}{n r^2} + \frac{\dot{r}'}{r'} - \frac{\dot{\Gamma}}{\Gamma} = 0.
\end{equation}
Now we impose the local conservation of energy, which in our case looks like
\begin{equation}
    \nabla_\mu T^{\mu\nu} = \alpha(\phi) T \nabla^\nu \phi.
\end{equation}
Taking the components parallel to \(u^\mu\) we get
\begin{equation}
    u_\mu \alpha(\phi) T \nabla^\mu \phi = \alpha(\phi) T \dot{\phi}
\end{equation}
for the RHS, while for the LHS we get the same expression as in GR, resulting in
\begin{equation}
    \dot{\rho} = \frac{{p} + \rho}{n} \dot{n} - \alpha(\phi) T \dot{\phi}. \label{eq_MS_FirstLaw}
\end{equation}
To obtain the Euler equation of the system, we study the components of the equation perpendicular to \(u^\mu\). For that, we define the projection \( P^{\alpha\beta} = g^{\alpha\beta} + u^{\alpha} u^{\beta} \), which leads us to
\begin{equation}
    P_{\alpha\beta} \alpha(\phi) {T} \nabla^\beta \phi = \alpha(\phi) T \left( \nabla_\alpha \phi + u_\alpha \dot{\phi} \right).
\end{equation}
Only the component $\alpha=1$ is non-trivial, which results in the Euler equation of the the fluid with an extra term to account for the interaction with the scalar field,
\begin{equation}
    p' + ({p} + \rho) \Phi' + \alpha(\phi) T \phi' \frac{r'^2}{\Gamma^2} = 0.
    \label{eq_MS_Euler}
\end{equation}

We must also ensure that the Einstein equations are satisfied. Starting with the component \(G_{01}\), we obtain
\begin{equation}
    G_{01} = 8\pi T_{\phi,01} \Rightarrow  8 \pi\dot{\phi} \phi' \Gamma r = -2  \left( \dot{\Gamma} r' - \Gamma \Phi' \dot{r} \right).
\end{equation}
If we define $U = \dot{r}$, it necessarily follows that
\begin{equation}
    U \Phi'= \dot{r}' - U'   .
\end{equation}
Substituting this into the equation above yields:
\begin{align}
    \frac{U'}{r'} = -\frac{\dot{\Gamma}}{\Gamma} + \frac{\dot{r}'}{r'} - \frac{4\pi r \dot{\phi} \phi' }{  r'} \nonumber \\ \implies \frac{\dot{(nr^2)}}{nr^2}+\frac{U'}{r'}= - \frac{4 \pi r \dot{\phi} \phi' }{ r'}.
\label{eq_MS_n}
\end{align}
On the other hand, for the \(G_{00}\) component we have:
\begin{align}
    G_{00} =& 8 \pi T_{00} + 8 \pi T_{\phi 00} \Rightarrow \nonumber \\
    \Gamma^2 =& -2 \left( \frac{m + m_\phi}{r} \right) + U^2 + 1,
    \label{eq_MS_Gamma}
\end{align}
where $m$ and $m_\phi$ represent the energy/mass of the matter and the scalar field, whose definitions are
\begin{equation}
    \begin{cases}
        m = \int 4\pi r^2 \rho \, r' d\mu, \\
        m_\phi = \int 4 \pi r^2 \left( -p_\phi + \dot{\phi}^2 \right)  r' \, d\mu.
    \end{cases}
\end{equation}
Finally, for \(G_{11}\) we have
\begin{align}
    G_{11} = 8\pi T_{11} + 8 \pi T_{\phi 11} \Rightarrow 
    \dot{U} = -\frac{{p'}\Gamma^2}{r' (p + \rho)} \nonumber \\ - \frac{m + m_\phi}{r^2}  - 4 \pi r \left(  {p} +  p_\phi +  {\phi'}^2 \frac{{\Gamma}^2}{r'^2} \right).
\end{align}

These equations together with a barotropic equation of state \( {p} = p(\rho) \) determine the collapse. The complete set of equations can be easily consulted in Table \ref{tabla_MS}.

\begin{table*}[t]\centering 
	\begin{tabular}{|c|c|}
		\hline
		$\dot{r}=U$& Dynamic equation for $r$       \\[8pt] \hline
		$ \frac{\dot{(nr^2)}}{nr^2}+\frac{U'}{r'}\textcolor{cyan}{+\frac{ 4\pi r \dot{\phi} \phi'}{  r'}}=0 $ & Algebraic equation for $n$       \\[8pt] \hline
		$\dot{\rho} = \frac{{p} + \rho}{n} \dot{n} \textcolor{cyan}{- \alpha(\phi) T \dot{\phi}}$   & First law of thermodynamics       \\[8pt] \hline
		$ \dot{U} = -\frac{{p'}\Gamma^2}{r' (p + \rho)}  - \frac{m  \textcolor{cyan}{+m_\phi}}{r^2}  - 4\pi r \left(  {p} \textcolor{cyan}{+  p_\phi +  {\phi'}^2 \frac{{\Gamma}^2}{r'^2}} \right)$ & Equation of motion      \\[8pt] \hline
		$p' + ({p} + \rho) \Phi' \textcolor{cyan}{+ \alpha(\phi) T \phi' \frac{r'^2}{\Gamma^2}} = 0$    & Euler equation      \\ \hline
		$m = \int 4\pi r^2 \rho \, r' d\mu$& Enclosed energy of the fluid       \\[8pt] \hline
            $\textcolor{cyan}{m_\phi = \int 4 \pi r^2 \left( -p_\phi + \dot{\phi}^2 \right) r' \, d\mu}$& Enclosed energy of the scalar field \\[8pt] \hline
		$\Gamma^2 = -2 \left( \frac{m  \textcolor{cyan}{+m_\phi}}{r} \right) + U^2 + 1$ &      Algebraic equation for $\Gamma$    \\[8pt] \hline
		$p=p(n,\rho)$&  Equation of state     \\[8pt] \hline
            $\textcolor{cyan}{p_\phi = \frac{1}{2} \left( \dot{\phi}^2 - \phi'^2 \frac{\Gamma^2}{r_{,a}^2} \right) - V(\phi)}$&  Scalar field pressure       \\[8pt] \hline
            $\textcolor{cyan}{\Box \phi = - \alpha(\phi) T + \frac{\partial V(\phi)}{\partial \phi}}$&  Klein-Gordon equation      \\[8pt] \hline 
	\end{tabular}
	\caption{Misner-Sharp equations for the spherically symmetric collapse modified to include the coupled scalar field. The black terms are the same as in the GR case while the blue terms represent the scalar-tensor modifications.} \label{tabla_MS}
\end{table*}

\section{Equilibrium configuration}\label{sec3}
In order to try to run a numerical simulation of the collapse based on these equations, it is crucial to introduce proper initial conditions. These could be obtained by taking the hydrostatic solution of the system and drastically reducing the internal pressure. The first step towards this is finding the corresponding TOV equation \cite{TOV}, which can easily be done in this formulation. Indeed, taking $U=0$ for the equilibrium configuration, 
\begin{equation} \small
    0 = -\frac{p'}{r^2(\rho+p)}\,\Gamma^2 -\frac{m+m_\phi}{r^2}
    -4\pi r\Bigl( p + p_\phi +  \phi^2\frac{\Gamma^2}{r'^2}\Bigr).
\end{equation}
From which one may deduce the relation
\begin{equation}
\begin{aligned}
    \frac{dp}{dr} = \frac{p'}{\Gamma^2} = &-\frac{\rho+p}{\Gamma^2} 
    \Biggl[ \frac{m+m_\phi}{r^2} \\
    & + 4\pi r \left( p + p_\phi + \phi'^2 \frac{\Gamma^2}{r'^2} \right) \Biggr].
\end{aligned}
\end{equation}
Substituting the value for \(\Gamma^2\) and simplifying we get:
\begin{equation}
\begin{aligned}
\frac{dp}{dr} = & -\frac{m+m_\phi}{r^2}\rho \left(1 + \frac{p}{\rho}\right) \left[ 1 - \frac{2(m+m_\phi)}{r} \right]^{-1} \\
& \cdot \left[ 1 + \frac{4\pi r^3}{m+m_\phi} \left( p + p_\phi + \frac{\phi'^2 \Gamma^2}{r'^2} \right) \right],
\end{aligned}
\end{equation}
which is the modified TOV equation associated with our problem. It generalizes the GR case by including the effects of the pressure and energy of the scalar field.

\section{Homogeneous interior solution}
The simplest non-trivial solution of the system is a homogeneous cosmological-like spacetime. We may interpret this as the interior of a highly idealized star. This solution is easily recovered by imposing that the thermodynamical variables and the scalar field are independent of the position, while the metric functions  $\Gamma$ and $\Phi$ are independent of time.\par

Then, it follows from equation (\ref{eq_MS_Euler}) that $\Phi$ must be a constant that we will assume to be $0$ for simplicity. On the other hand, we may solve for $\Gamma$ by introducing the variable separation $r(\mu,t_c)=R(\mu)f(t_c)$ into equation (\ref{eq_MS_Gamma}):
\begin{equation}
    \Gamma^2=-2  \frac{m+m_\phi}{R(\mu) f(t_c)}+  R(\mu)^2 \dot{f}(t_c)^2 +1.
\end{equation}
Substituting the corresponding values for $m$ and $m_\phi$ in this expression and separating the terms depending on $\mu$ from the ones that depend on  $t_c$, we can define the constant $k$ as
\begin{equation}
\begin{split}
    \frac{\Gamma^2-1}{R(\mu)^2} &= \dot{f}(t_c)^2 - \frac{8 \pi}{3} f(t_c)^2 \\
    & \cdot \left( \rho + \frac{1}{2} \dot{\phi}^2 + V(\phi) \right) = -k.
\end{split}
\label{eq_separation_FLRW}
\end{equation}
The metric then takes the form 
\begin{equation}
    ds^2=-dt^2+f(t_c)^2\left( \frac{1}{1-kR^2}dR^2+R^2d\Omega^2 \right),
\end{equation}
which is the usual FLRW metric. In fact, the other side of equation (\ref{eq_separation_FLRW}) leads to the corresponding Friedman equation:
\begin{equation}
    \frac{\dot{f}(t_c)^2}{f(t_c)^2}+\frac{k}{f(t_c)^2}-8 \pi \left(  \rho+\frac{1}{2} \dot{\phi}^2 + V(\phi) \right)=0. \label{eq_Friedmann_FLRW}
\end{equation}
The concrete value of $k$ determines the curvature in the interior of the star and might depend on the initial configuration of the system, as is the case in the Oppenheimmer-Snyder collapse. The Klein-Gordon equation (\ref{eq_KG}) that dictates the evolution of the scalar field in this spacetime takes the form
\begin{equation}
    \ddot\phi+3 \frac{\dot{f}}{f} \dot{\phi} +\frac{\partial V( \phi )}{\partial \phi}= \alpha ( \phi )T. \label{eq_KG_FLRW}
\end{equation}

If we now restrict ourselves to the Brans-Dicke theory, in which $\alpha$ is constant and there is no potential, and assume the fluid is composed of dust particles (null pressure), we may obtain the following continuity equation from (\ref{eq_MS_FirstLaw}) and (\ref{eq_MS_n}),
\begin{equation}
    \dot{\rho}+3\frac{\dot{f}}{f}\rho=-\alpha \rho \dot{\phi} \label{eq_continuity_FLRW}.
\end{equation}
Both equations, (\ref{eq_KG_FLRW}) and (\ref{eq_continuity_FLRW}), are equal to the ones found in a cosmological scenario with matter and a scalar field except for the inclusion of the coupling term between the two. It is possible to explicitly solve the matter continuity equation by integrating with respect to time:
\begin{equation}
    \rho=\rho_{i}f^{-3}e^{-\alpha(\phi-\phi_{i})}.
\end{equation}
The result is the usual evolution for the energy density of matter weighted by a function of the scalar field, with the $i$ subindex indicating that the function is evaluated at the initial time $t_c=t_i$.

\begin{figure}[h]
\includegraphics[width=1.1\columnwidth]{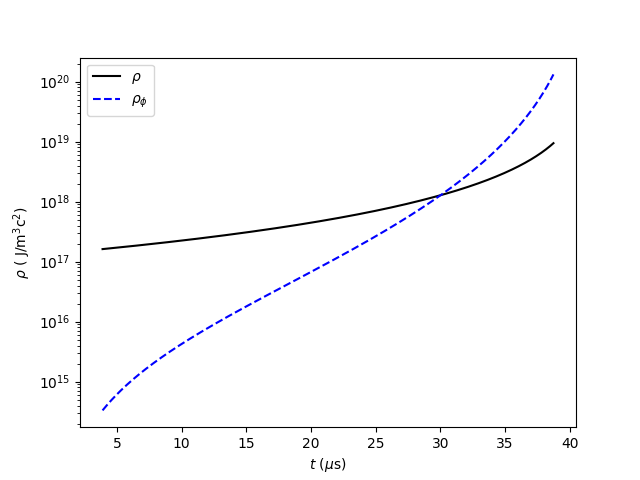}
	\caption{Evolution in time of the matter energy density (black) and the scalar field energy density (blue) for an idealized homogeneous stellar collapse choosing $\alpha=0.2$.}
	\label{Fig:FLRW_rho}
\end{figure}

\begin{figure}[h]
\includegraphics[width=1.1\columnwidth]{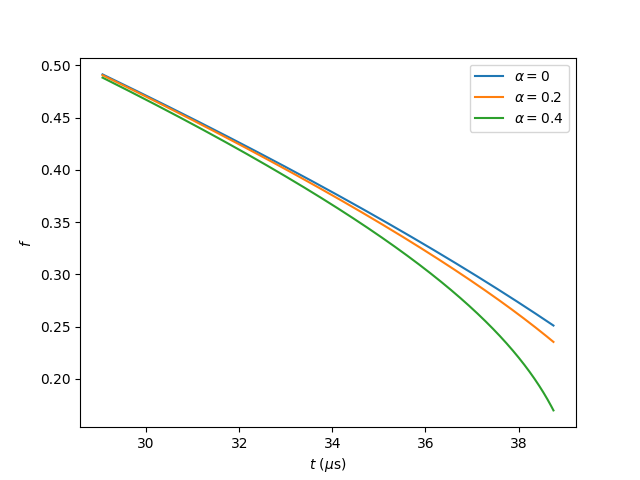}
	\caption{Evolution of the scale factor $f$ in a homogeneous stellar collapse for different values of the coupling constant $\alpha$. The line at the top corresponds to $\alpha=0$ (blue), the middle one to $\alpha=0.2$ (orange), and the one in the bottom to $\alpha=0.4$ (green).
    }
	\label{Fig:FLRW_f}
\end{figure}

Since in order to match our observations the theory must be close to GR \cite{Amirhashchi2019}, it is safe to assume that the coupling constant $\alpha$ is close to $0$. This allows us to solve perturbatively for $\phi$:
\begin{equation}
    \phi \approx \phi_i+8 \pi \alpha f_i^3 \rho_i \int^{t_{c}}_{t_i}\frac{t_1}{f(t_1)^3}dt_1. \label{eq_phi_FLRW}
\end{equation}
Which indicates that the interior scalar field will grow as the star collapses. Alternatively, one could try to numerically solve the system formed by equations (\ref{eq_Friedmann_FLRW}) and (\ref{eq_KG_FLRW}) and use the solutions to compute $\rho$. This was done for Figures \ref{Fig:FLRW_rho} and \ref{Fig:FLRW_f} using a simple finite differences method and setting  $k=0$. What we see is that, while originally the energy of the scalar field is negligible, it grows significantly faster than its matter counterpart, becoming dominant in the latter stages of the collapse. Complementing this information, we also notice that while the value of $\alpha$ does not seem to have an impact in the initial phases, the collapse is drastically accelerated for a stronger coupling constant later on.

The homogeneous interior serves as an idealized framework designed to capture the fundamental qualitative dynamics of the collapse. While this scenario simplifies the internal structure, it provides a clear analytical window into the underlying physical mechanisms that might be obscured in more complex configurations. Although a perturbative analysis—analogous to those in cosmological contexts \cite{Weinberg}—indicates that density perturbations grow faster than the background density during contraction, suggesting an inherent instability, there is strong evidence that the global features of the collapse remain qualitatively robust. Indeed, established studies \cite{MW, VR}, even when considering coupled scalar fields \cite{Price72}, support the view that the general evolutionary features are not fundamentally altered by the presence of inhomogeneities. Consequently, we adopt this treatment as a consistent and tractable starting point for our analysis.

\section{Exterior region}\label{Exterior region}
One has to be careful when choosing the spacetime for the region surrounding the star. Unlike in standard GR, where one can simply choose the spacetime to be static, the inclusion of a scalar field leads to the emission of radiation even when spherical symmetry is imposed. A numerical analysis of the scalar radiation emitted by a gravitational collapse similar to the one considered here can be found in \cite{Harada97} and \cite{Shibata94}.
Despite these considerations, the assumption of a static exterior metric remains a consistent approximation during the early stages of the collapse. Given that both the star and its surrounding spacetime are assumed to be initially at rest, this staticity holds while the stellar radius is sufficiently large and the dynamical evolution remains slow.\par 
The only static and spherically symmetric solution of the Einstein equations for a non-trivial massless scalar field is known to be the JNW metric \cite{JNW68}\footnote{The new coordinates $t$ and $r$ shall not be confused with the coordinate $t_c$ and the function $r(t_c,\mu)$ that were introduced previously.},
\begin{multline}
    ds^2 = -\left(1 - \frac{b}{r}\right)^{\beta} dt^2 
+ \left(1 - \frac{b}{r}\right)^{-\beta} dr^2  \\
+ \left(1 - \frac{b}{r}\right)^{1 - \beta} r^2 d\Omega^2.
\end{multline}
Where $b=2M/\beta$ is an effective mass and the parameter $\beta$ relates to $\alpha$ via $\beta=\sqrt{1-\alpha^2}$. In this solution, the scalar field must take the form
\begin{equation}
    \phi=-\frac{\alpha}{\sqrt{2}}\ln{\left[{\left( 1-\frac{b}{r}\right)^\beta}\right]}.
\end{equation}

While formally similar, the causal structure of this spacetime is essentially different to that of Schwarzschild. This is because the horizon, $r_g=2b$, is not a coordinate singularity but a true naked singularity causally connected to every point in the manifold. This means that even after taking the limit $\alpha \rightarrow 0$ one would not recover the complete Schwarzschild but a version of it truncated at the event horizon (see \cite{Matsuda71}). Hence, by taking this to be the exterior during the entirety of the collapse, the resulting spacetime must contain a naked singularity. However, because we are only using it as an approximation for the initial stages, we shall not arrive at any conclusion about the fate of the collapse. To describe the exterior region at later stages of the collapse, one should ideally employ a spacetime capable of capturing the gravitational effects of radiation. A detailed analysis on how to derive such solutions within the Jordan frame can be found in \cite{Kaufmann68}.\par

Let us now notice that since the metric does not explicitly depend on the coordinate $t$, $\xi^{\mu}=(1;0,0,0)$ is a killing vector with an associated conserved quantity $E$, which for a particle of four-momentum $p^\mu=u^\mu m_0e^{\phi(r)}$\footnote{In the Einstein frame the effective mass is modified by the interaction with the scalar field. Hence, the inclusion of the exponential factor in the four-momentum of the particle.}, one may derive the expression\footnote{We have absorbed $m_0$ into $E$ without loss of generality.}
\begin{equation}
   p^\mu\xi^\nu g_{\mu \nu}=-E \implies \frac{dt}{d\tau}=\frac{E e^{\phi/(\alpha\beta^2)}}{A(r)^{\beta(\sqrt{2}/2+1)}},\label{eq_defE}
\end{equation}
in which $A(r)=\left( 1-b/r\right)$ and $\tau$ is the proper time of the particle. This quantity is identified as the generalized energy of the particle, and it can be proven that it is strictly conserved, even when considering the influence of the scalar field on the particle trajectory.
If the particle is radially falling towards the center, the corresponding 4-velocity must then satisfy
\begin{multline}
    -1=u^\alpha u^\beta g_{\alpha \beta} \\ \implies E^2e^{\phi/(\alpha \beta^2)}A^{-\sqrt{2}\beta}=\dot{r}^2+A(r)^\beta.
    \label{eq_radial}
\end{multline}
Assuming the particle was originally at rest and began to fall from an initial radius $r_i$, we find that the energy is determined by the initial radius via
\begin{equation}
    E^2e^{\phi(r_i)/(\alpha \beta^2)}=A(r_i)^{\beta(\sqrt{2}+1)}.
\end{equation}

Since $r(\tau)$ must go from $r_i$ at the beginning of the collapse and becomes progressively smaller as it continues, we may introduce the following parametrization for $r$,
\begin{equation}
    r(\eta)=\frac{r_i}{2}(1+\cos{\eta}).
\end{equation}
Substituting this into equation (\ref{eq_radial}), ignoring the terms of order $O(1/r^2)$ and integrating, it follows that the value of $\tau$ in terms of the new parameter $\eta$ must be
\begin{equation}
    \tau(\eta)=\frac{r_i}{4}\left( \frac{M(\sqrt{2}\alpha^2-1)}{2\sqrt{2} M \alpha^2-r_i}\right)^{-1/2}(\eta +\sin{\eta}).
\end{equation}
These are the equations of a cycloid and are formally similar to the ones found in the Oppenheimer-Snyder scenario. Therefore, we can conclude that, at the beginning of the collapse, the scalar field does not radically alter the motion of the star surface.

\section{Junction conditions}

So far we have deduced a cosmological spacetime for the interior of the star and have imposed a static solution for the exterior, but it remains to be proven whether this combination actually constitutes a physically reasonable spacetime describing the collapse. In order for this to be the case, we must make sure our solution satisfy the junction conditions at the surface of the star (see \cite{Israel66,Poisson}).\par

Before this, we must properly define the hypersurface $S$ that will serve as a separation between the two spacetimes. For the interior region this is quite straight forward, since $R$ is comoving to the fluid, we can simply consider a spherical shell $R=R_s$ to be the surface of the star. The normal vector to this surface is
\begin{equation}
    n_-^\mu=\left( 0,\sqrt{\frac{1-kR_s^2}{f(t)^2}},0,0\right).
\end{equation}
While not as analytically simple, the surface on the exterior region is easily defined by taking $t$ and $r$ to be functions of $\tau$ that follow equations (\ref{eq_defE}) and (\ref{eq_radial}) respectively. In this case, the normal vector is found by asking it to be orthogonal to the angular components and to the 4-velocity of the surface,
\begin{equation}
    n_+^\mu u_\mu=n_+^\theta=n_+^\varphi=0,
\end{equation}
and properly normalizing it so that it corresponds to a timelike surface
\begin{equation}
    n_+^\mu n_+^{\nu}g_{\mu \nu}=1.
\end{equation}

The first junction condition consists on making sure the induced metric over the hypersurface is equal for both regions. The most efficient way of doing this is by working with a reduced set of coordinates that can easily describe any point on the hypersurface. In our case the obvious choice is using the coordinates $\phi$ and $\theta$ together with $\tau$, the proper time of a particle on the star surface. Notice that since the cosmological coordinate $t_c$ already represents the proper time of the fluid elements of the spacetime, the identity $t_c=\tau$ becomes trivial. With this in mind, it follows that the induced interior metric takes the form
\begin{equation}
    ds^2_{-}|_{S}=-d\tau^2+R_s^2f(\tau)^2d\Omega^2.
\end{equation}
On the other hand, for the exterior metric, we have,
\begin{equation}
    ds^2_{+}|_{S}=-d\tau^2+A \left( r(\tau) \right)^{1-\beta}d\Omega^2.
\end{equation}
Therefore, the first junction condition will be satisfied if and only if the following equation is true:
\begin{equation}
    R_s^2f(\tau)^2=A(r(\tau))^{1-\beta}r(\tau)^2.
    \label{eq_cond1}
\end{equation}

Setting $\alpha=0$ recovers the GR scenario, where the equation can be proven to hold relatively simply. For higher orders of $\alpha$, the function $f(\tau)$ becomes increasingly complicated, and an analytical treatment of the problem is no longer feasible. Nevertheless, we conjecture that the equality remains valid under the approximations considered. This conjecture is supported by two key arguments: the JNW metric approaches the Schwarzschild solution for sufficiently large $r$, and the parameter $\alpha$ is known to be small.

For the second junction condition, we ask the extrinsic curvature to be the same on both sides of the hypersurface. Using the same coordinates that we used for the induced metric, the extrinsic curvature for the interior will be

\begin{equation}
    K^-_{\theta \theta}=K^-_{\varphi \varphi}/\sin^2{\theta}=R_sf(\tau) \sqrt{-R_s^2 k +1},
\end{equation}
with any other term being null. While for the exterior, the non-null terms of the extrinsic curvature are
\begin{equation}
    K^+_{\tau \tau}=\frac{\sqrt{2}EM(1-\beta^2)e^{\phi/(\alpha \beta^2)}A(r(\tau))^{-\sqrt{2}\beta/2}}{(2M-\beta r(\tau))r(\tau)},
\end{equation}
\begin{multline}
    K^+_{\theta \theta}=K^+_{\varphi \varphi}/\sin^2{\theta}=
    Ee^{\phi/(\alpha \beta^2)}A(r(\tau))^{-\beta(1-\sqrt{2})} \\ \cdot  \left( -M -M/\beta + r(\tau)\right).
\end{multline}
The first of these terms is of order $O(1/r^2)$ and therefore vanishes under our approximations. Note that in both cases, $K^+$ and $K^{-}$, we have $K_{\varphi \varphi}=\sin^2{\theta K_{\theta \theta}}$, meaning that our second condition reduces to only one equation. After neglecting terms of order $O(1/r)$ and substituting the value for $f(\tau)$ from equation (\ref{eq_cond1}) we get
 \begin{equation}
    E^2=1-kR_s^2,
\end{equation}
a relation between the constants of the inside and outside geometry which is also found in the Oppenheimer-Snyder collapse.

Beyond the usual junction conditions for the metric, the scalar field must also satisfy certain conditions in the frontier between the two regions (see, for instance, \cite{Sakai93}). Namely, one must guarantee the continuity of the scalar field,
\begin{equation}
    \phi^+=\phi^-,
    \label{eq_phi_cond1}
\end{equation}
and that of its derivative in the direction normal to he hypersurface,
\begin{equation}
    n_-^\mu \nabla_\mu \phi ^-= n_+^\mu \nabla_\mu \phi ^+.\label{eq_phi_cond2}
\end{equation}

Similarly to what happened with the first condition of the metric, when analyzing the continuity of the scalar field the problem becomes analytically intractable. Despite this, it is possible to say that for small enough values of $\tau$ the relation (\ref{eq_phi_cond1}) holds if one sets
\begin{equation}\label{eq_phi_small_t}
    \phi_i=\sqrt{2}\alpha(1-E^2).
\end{equation}

Moving into the continuity of the derivative, it is easy to check that it is null for the interior metric,
\begin{equation}
    n_-^\mu \nabla_\mu \phi ^-=0,
\end{equation}
and that despite its apparent complexity, this term is suppressed in the exterior region at large distances. Specifically, by considering only the leading-order terms in the $1/r$ expansion, we find,
\begin{equation}
    n_+^\mu \nabla_\mu \phi ^+=\frac{\sqrt{2}EM\alpha e^{\frac{2}{\sqrt{2}}\alpha^2 \ln{\left[ A(r)^\beta\right]}}}{r(b-r)}=\mathcal{O} \left(\frac{1}{r^2}\right).
\end{equation}
Consequently, equation (\ref{eq_phi_cond2}) is satisfied, confirming that the smooth matching of the two spacetimes is consistent within the adopted approximations. This validates the choice of a static exterior metric, provided the second-order terms in $1/r$ remain negligible and the time elapsed since the onset of collapse is sufficiently small for Eq. (\ref{eq_phi_small_t}) to hold.

\section{Discussion}
In this work, we have provided a systematic and comprehensive framework for studying spherically symmetric gravitational collapse within the context of a general scalar-tensor theory of gravity, formulated entirely in the Einstein frame. Our approach aimed to bridge the gap between analytical strong-field solutions and the robust numerical methods required to model realistic astrophysical scenarios.

Our primary theoretical contribution is the derivation of a complete set of modified Misner–Sharp equations for scalar-tensor gravity. This generalized formulation, which includes new terms representing the contributions of the scalar field on the fluid dynamics, is fully prepared for numerical implementation, providing a crucial tool for future simulations of collapse processes. 

Building upon this, we obtained the corresponding modified TOV equation that describes the equilibrium state of a static star in the presence of the scalar field. This modified equilibrium condition is essential for establishing the initial data for any collapse simulation, as it explicitly determines how the scalar field alters the fundamental balance of forces (pressure gradients vs. gravity) and thus influences the maximum mass and stability of compact objects in these theories. To gain physical insight into the early stages of collapse, we analyzed an instructive model based on a homogeneous dust star (FLRW interior) matching a static exterior spacetime. The successful and consistent junction with the JNW metric suggests that this approximation is a proper description for the initial stages of the process. 

This model allowed us to analytically demonstrate several key features. For instance, the coupling accelerates the overall collapse process compared to standard GR. In fact, the energy density of the scalar field can grow rapidly, dominating the total energy density of the fluid as the collapse proceeds, indicating the critical role of the scalar dynamics even in the initial phases. The rigorous approach to the interior-exterior matching stands as an independent verification of the theoretical consistency of the initial collapse phase in these theories, establishing the necessary conditions on both the metric and the scalar field at the stellar surface. 

Our findings, particularly the derived Misner–Sharp equations and the collapse analysis, reinforce the understanding that the scalar degree of freedom fundamentally alters the gravitational collapse process. The literature presents a persistent debate regarding the final state of gravitational collapse in scalar-tensor theories. Naively, one might anticipate the formation of a compact object whose exterior is described by the naked singularity present in the JNW metric, suggesting a potential pathology in the theory. However, theoretical arguments \cite{Penrose,Thorne71,Hawking72} and advanced numerical relativity simulations \cite{Scheel94a,Scheel94b,Shibata94,Novak98} have indicated that, depending on the scalar model and under astrophysically realistic conditions, a black hole identical to the one found in standard GR could be formed, and the scalar field is effectively expelled. Recent theoretical work continues to support this conclusion \cite{Ziaie2022}.

In spite of these significant advancements, several crucial questions remain open: The precise mechanism by which the scalar field {shuts off} or becomes constant during the dynamic collapse process, satisfying the no-hair conjecture, is still not fully understood across all scalar-tensor models. Indeed, most simulations have focused on the classic Brans–Dicke theory. How the collapse evolves in more general scalar-tensor scenarios, which include arbitrary potentials or more complex couplings, remains largely unexplored. While the present work did not directly address these final-state questions, the new framework developed here is designed to facilitate their exploration. The generalized Misner-Sharp formulation is readily extendable to incorporate different coupling functions, nontrivial potentials, or more complex equations of state. We believe that this formalism, coupled with the insight gained from our analysis of the initial phases, provides the essential analytical and computational foundation for future work aiming to resolve the outstanding issues in black hole formation and to constrain modified gravity theories through astrophysical observations.

Alternatively, recent studies have explored the possibility that gravitational collapse may result in exotic singularity-free objects, such as gravastars \cite{Mazur, Sinha1, Sinha2, Sinha3, Sinha4}, rather than traditional black holes. Although the precise formation mechanisms for these objects remain an open question, significant progress has been made in providing theoretical frameworks for their existence \cite{Rezzolla25}.

In summary, this work provides a comprehensive analytical and computational framework for studying gravitational collapse in scalar-tensor gravity, combining modified hydrodynamics, equilibrium solutions, interior-exterior matching, and an illustrative example to elucidate the complex role of scalar interactions in strong-field astrophysics. The derived equations represent a powerful tool for numerical modelers seeking to explore the gravitational waveforms and potential observable deviations from GR that may arise from these theories.

\bmhead{Acknowledgements}

 This work is supported by the project PID2022-139841NB-I00 funded by MICIU/AEI/10.13039/501100011033 and by ERDF/EU. This work is also part of the COST (European Cooperation in Science and Technology) Actions CA21106, CA21136, CA22113 and CA23130.

\end{document}